\documentclass[12pt]{iopart}

\usepackage{graphicx}
\usepackage[utf8x]{inputenc}
\usepackage{hyperref}
\usepackage{todonotes}
\usepackage{iopams}
\usepackage{amssymb}
\expandafter\let\csname equation*\endcsname\relax
\expandafter\let\csname endequation*\endcsname\relax
\usepackage{amsmath}
\usepackage{upgreek}
\usepackage{fixmath}
\usepackage{lmodern}
\usepackage{subfig}
\usepackage[square, sort&compress]{natbib}
\bibpunct{[}{]}{,}{n}{}{}
\usepackage{soul}
\definecolor{lightblue}{rgb}{.90,.95,1}
\sethlcolor{lightblue}

\renewcommand{\epsilon}{\varepsilon}
\renewcommand{\vec}[1]{\ensuremath{\boldsymbol{#1}}} 

\newcommand{\grad}{\nabla}
\newcommand{\Dd}{\ensuremath{\mathrm{d}}}

\newcommand{\unit}[1]{\ensuremath{\,\mathrm{#1}}}
\newcommand{\D}[2]{\ensuremath{\frac{\partial#2}{\partial#1}}}

\newcommand{\Eqref}[1]{equation~\eqref{#1}}

\newcommand{\Figref}[1]{figure~\ref{#1}}

\newcommand{\Tabref}[1]{table.~\ref{#1}}

\begin{document}


\title{Effects of the equilibrium model on impurity transport in tokamaks}
\author{A~Skyman, L~Fazendeiro, D~Tegnered, H~Nordman, J~Anderson and P~Strand}
\address{Department~of~Earth~and~Space~Sciences,
         Chalmers~University~of~Technology, SE-412~96~Gothenburg, Sweden}
\eads{\mailto{andreas.skyman@chalmers.se}, \mailto{luisfa@chalmers.se}}

\begin{abstract}
Gyrokinetic simulations of ion temperature gradient mode and trapped electron mode driven impurity transport in realistic
tokamak geometry are presented and compared with results using simplified
geometries.
The gyrokinetic results, obtained with the GENE code in both linear and
non-linear mode, are compared with data and analysis for a dedicated impurity injection
discharge at JET.
The impact of several factors on heat and particle transport are discussed,
lending special focus to tokamak geometry and rotational shear.
To this end, results using $s-\alpha$ and concentric circular equilibria are
compared to results with magnetic geometry from a JET-experiment.
To further approach experimental conditions, non-linear gyrokinetic simulations
are performed with collisions and a carbon background included.

The impurity peaking factors, computed by finding local density gradients
corresponding to zero particle flux, are discussed.
The impurity peaking factors are seen to be reduced by a factor
of $\sim 2$ in realistic geometry compared to the simplified geometries,
due to a reduction of the convective pinch.
It is also seen that collisions reduce the peaking factor for low $Z$
impurities, while increasing it for high charge numbers, which is attributed to
a shift in the transport spectra towards higher wave-numbers with the addition
of collisions.
With the addition of roto-diffusion, an over-all reduction of the peaking
factors is observed, but this decrease is not sufficient to explain the flat
carbon profiles seen at JET.
\end{abstract}

\maketitle


\section{Introduction}
\label{sec:intro}
Impurity transport is a matter of crucial relevance for Tokamak fusion
plasmas~\cite{Frojdh1992, Basu2003, Estrada-Mila2005, Naulin2005, Priego2005,
Fulop2006, Bourdelle2007, Dubuit2007, Camenen2009, Fulop2010, Futatani2010,
Hein2010, Moradi2010, Fulop2011, Nordman2008, Angioni2006, Angioni2007,
Nordman2007a, Angioni2009a, Fulop2009, Nordman2011, Skyman2011a, Skyman2011b,
Mollen2012, Moradi2012, Kazakov2013, Mollen2013, Dux2003, Puiatti2003,
Puiatti2006, Giroud2007, Giroud2009} due to the contribution of impurities to
radiation losses and plasma fuel dilution.
The impurities can originate either from the sputtering of wall and divertor
materials or from deliberate impurity injection in order to reduce power load.
The impurities in a tokamak thus cover a large range in charge number $Z$,
making it necessary to study the scaling of impurity transport with $Z$.
This is even more relevant after the installation of the new ITER-like wall at
JET~\cite{Matthews2009}, with its beryllium wall and tungsten divertor.
In addition, a typical plasma discharge already has such a low particle
density, that even a modest amount of impurities can greatly dilute the plasma,
thereby reducing the power output.
Since most impurity sources are located at the edge of the tokamak, the
resulting impurity profile in the core of a given discharge will depend on the
balance between diffusive and advective processes.

It is well known empirically that shaping of the plasma has a beneficial effect
on energy confinement, the Troyon limit and the Greenwald density limit.
For example, elongation ($\kappa$) is one of the factors considered in
empirical scaling laws (e.g.~\cite{IPB2}) of the energy confinement time.
Only in recent years, however, has the impact of the geometric configuration on
micro-turbulence been rigorously ascertained, e.g. from simulations of the
gyrokinetic equations~\cite{Xanthopolous2006, Belli2008, Lapillonne2009,
Burckel2010, Told2010}.

It has been observed that the heat-flux scales with elongation at fixed average
temperature gradients, and that shaping effects enhance residual zonal flows,
thus increasing the nonlinear critical temperature gradient~\cite{Belli2008}.
Elongation has been seen to be the dominant effect on micro-turbulence on
ion temperature gradient~(ITG) and trapped electron~(TE) drift-wave instabilities, compared with e.g. triangularity, up--down
asymmetry and Dimits shift~\cite{Burckel2010}.
However, in spite of these results, the effects of the equilibrium model on
impurity transport has not been studied in any detail.

Previous work in this area includes~\cite{Nordman2011, Skyman2011a,
Skyman2011b}, in which the impurity transport in a dedicated impurity injection
experiment at JET was studied using $s-\alpha$ geometry.
It was shown that the profile steepness parameter of the injected impurities
was qualitatively reproduced by the simulations.
On the other hand, a discrepancy was found for the background carbon profile
which was too steep at mid radius compared to the experimental profile.

In the present study, an experimental equilibrium is used and compared with
results using simpler $s-\alpha$ and circular equilibria~\cite{Lapillonne2009}.
Since the impurity profile is a result of a balance between diffusion and
advection, where the main advective term is the inward curvature pinch, a
strong dependence on the equilibrium model may be expected.
In addition, a more realistic physics description, including
the effects of rotational shear on the instabilities and impurity
transport (roto-diffusion), is investigated using gyrokinetic simulations.
Realistic non-linear gyrokinetic simulations are performed taking
into account collisionality, and the presence of a $2\%$C background,
consistent with the conditions applying for an impurity injection experiment at
JET.


The paper is organised as follows.
Section~\ref{sec:bg} describes the gyrokinetic simulations performed,
and the fundamental concepts of impurity transport.
Section~\ref{sec:profiles} describes the particular JET discharge considered.
In Section~\ref{sec:results}, results for the the eigenvalues, turbulent
fluctuation levels and transport are presented, along with a discussion on the
scaling of the impurity peaking factor with impurity charge $Z$ for the various
models considered.
A conclusion then follows in Section~\ref{sec:conclusions}.

\section{Background}
\label{sec:bg}

    \subsection{Gyrokinetic model}
    \label{sec:gyro}
The gyrokinetic simulations in this study were performed using the gyrokinetic
turbulence code
GENE\footnote{\url{http://www.ipp.mpg.de/~fsj/gene/}}~\cite{Jenko2000,Merz2008a}
where a flux--tube domain was considered.
This is an Eulerian--type code (i.e., employing a fixed grid in phase space),
which allows for arbitrarily long simulation times.
Both quasi- and nonlinear simulations including kinetic ions and electrons, and
at least one impurity species were performed.
All impurities were treated as fully kinetic species with low concentrations.
For studying the effects of collisions, we considered the GENE implementation
of the Landau--Boltzmann type collision operator~\cite{Merz2008a}.

For the simulation domain, all three geometries used a flux tube with periodic
boundary conditions in the perpendicular plane.
The nonlinear simulations were performed using a $96\times 96\times 32$ grid in
the normal, bi-normal, and parallel spatial directions respectively; in the
perpendicular and parallel momentum directions, a $12\times 48$ grid was used.
For the linear and quasilinear computations, a typical resolution was $12\times
32$ grid points in the normal and parallel directions, with $12\times 64$ grid
points in momentum space.
To validate the choice of resolutions, convergence tests for mode structure
and spectra were performed.

The nonlinear simulations were typically run up to $t=600\, a/c_\text{s}$ for
the experimental geometry scenario, where $a$ represents minor radius and
$c_\text{s}=\sqrt{T_e/m_i}$.
For the ``full scenario,'' in which a $2\%$C background and collisions were
both present, the simulation time was instead $t \approx 450\, a/c_\text{s}$.
This shorter time span is due to the much more computationally intensive nature
of the second scenario, by a factor of $\sim 2$--$3$.
The increased numerical cost was mainly due to collisions, though in part also
due to the addition of the $2\%$C background.

    \subsection{Experimental profiles and parameters}
    \label{sec:profiles}
\begin{figure}
    \centering
    \subfloat[profiles of $T_{i,e}$ and $n_i$]
             {\includegraphics{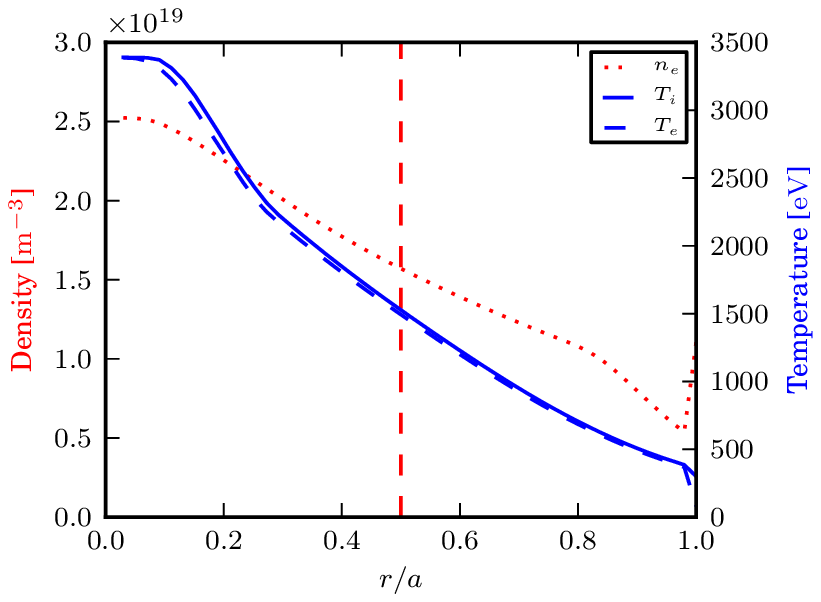}\label{fig:profiles}}~
    \subfloat[profiles of $v_\text{tor}$, $q$ and $s$]
             {\includegraphics{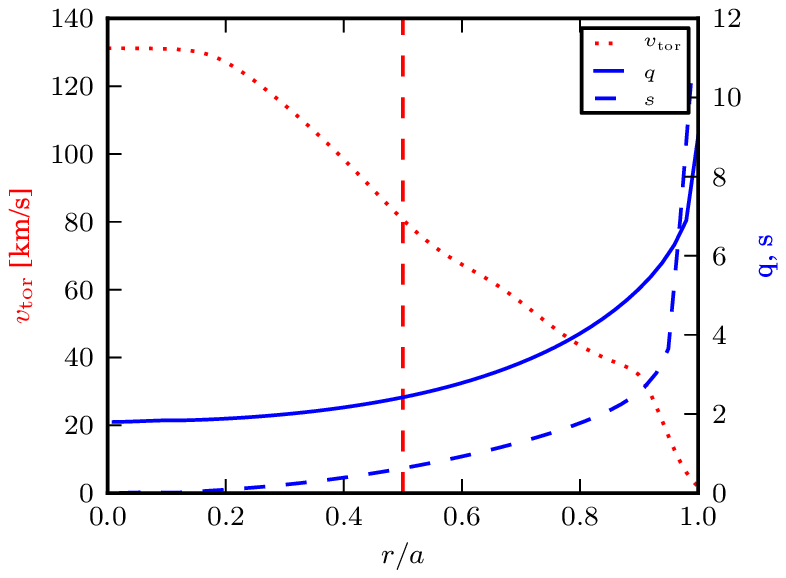}\label{fig:vtor}}~
    \caption{(colour online)
             Radial profiles of the background parameters for JET discharge
             \#67730 at $t=47.5\unit{s}$. The simulations were performed at mid
             radius ($r/a=0.5$; indicated).  }
\end{figure}

The physical parameters used in the simulations are presented in
\Tabref{tab:parameters} and were chosen so as to be consistent with the
experimental values taken from the JET~database for discharge \#67330.
This was previously analysed in~\cite{Nordman2011}, where the $s-\alpha$
geometry model was used.
The discharge was part of an impurity-dedicated set of
discharges with low MHD activity~\cite{Giroud2009}.
In these, extrinsic impurities (Ne, Ar and Ni) were injected via laser ablation
and gas injection.
Diffusivity $D_Z$ and convective velocity $V_Z$ were determined by matching
spectroscopic data with results obtained from predictive transport codes (see
\cite{Nordman2011} and in particular Fig.~4).
Profiles of density and temperature for the discharge are shown in
\Figref{fig:profiles}.

For the present study, the parameters used, including the MHD equilibrium (from
now on referred to as ``experimental geometry''), were taken extracted at
mid radius ($r/a=0.5$) at $48\unit{s}$.
The main quantities of interest were
elongation $\kappa=1.37$, triangularity $\delta=0.044$, and the fraction of
trapped particles $f_t = 0.55$.
The toroidal magnetic field intensity on the axis was $B=3\unit{T}$, the major
radius of the tokamak $R = 3\unit{m}$, electron temperature $T_e =
1.55\unit{k}\unit{eV}$ and plasma $\beta=8\pi n_e T_e/B^2=1.28 \times 10^{-3}$,
and the normalised gradient of $\beta$ $\alpha=-q^2 R \Dd \beta/\Dd r =0.1026$.
The safety factor in the discharge was $q=2.2$ at the considered radius,
obtained from EFIT.
The collisionality was $\nu_{ei} = 0.07\,c_\text{s}/R$
($\nu_c=0.28\times10^{-3}$ in GENE).
Effects of purely toroidal rotation were included through the
$\vec{E}\times\vec{B}$ shearing rate.
The normalised gradient scale lengths are defined as $R/L_{n_j}= -(R/n_j)(\Dd
n_j / \Dd r)$ and $R/L_{T_j} = -(R/T_j)(\Dd T_j / \Dd r)$ where $R$ is the
major radius of the tokamak.
For the realistic equilibria, the gradients were rescaled to the minor radius
with $a=\sqrt{2 \Phi_\text{s}/B}=1.26\unit{m}$, where $\Phi_\text{s}$ is the
toroidal flux at the last closed flux surface.
In the considered discharge, the Mach number was $M=0.21$, leading to
$\gamma_{\vec{E}\times\vec{B}} = -\frac{r}{q}\frac{1}{a}\D{r}{v_\text{tor}} =
0.1\, c_\text{s}/R$, with the toroidal velocity profile ($v_\text{tor}(r)$) as
in \Figref{fig:vtor}, evaluated at mid radius.
Hence, we only consider the effect of flow shear in the limit where the flow is
small, neglecting effects of centrifugal and Coriolis forces.
These may, however, be important for heavier impurities~\cite{Camenen2009}.

\begin{table}[tb]
    \centering
    \caption[Parameters]{\small Parameters for gyrokinetic simulations}
    \label{tab:parameters}
    \begin{tabular}{l r}
        \hline\hline
        $T_i/T_e$:                          & $1.02$ \\
        $\beta$:                            & $1.28\times10^{-3}$ \\
        $s$:                                & $0.75$ \\
        $q$:                                & $2.20$ \\
        $\epsilon=r/R$:                     & $0.17$ \\
        $\alpha$:                           & $0.126$ \\
        $\kappa$:                           & $1.37$ \\
        $\delta$:                           & $0.044$ \\
        $k_\theta \rho_\text{s}$:           & $0.2$--$0.6$ \\
        $n_e$, $n_i + Z\,n_Z$:              & $1.0 \cdot 10^{19} \unit{m^-3}$ \\
        $n_Z$ \emph{(trace)}:               & $10^{-6} \cdot 10^{19} \unit{m^-3}$ \\
        $Z$:                                & $2$--$74$ \\
        $R/L_{n_{i,e}}$:                    & $2.7$ \\
        $R/L_{T_i},R/L_{T_Z}$:              & $5.6$ \\
        $R/L_{T_e}$:                        & $5.6$ \\
        $\nu_c$:                            & $0$, $0.28\times10^{-3}$ \\
        $\gamma_{\vec{E}\times\vec{B}}$:    & $0$--$0.6$ \\
        $T_e$:                              & $1.55\unit{k}\unit{eV}$ \\
        $B$:                                & $3.04\unit{T}$ \\
        \hline\hline
    \end{tabular}
\end{table}

\subsection{Magnetic equilibrium}
\label{sec:geometry}
\begin{figure}
    \centering
    \includegraphics[height=60mm]{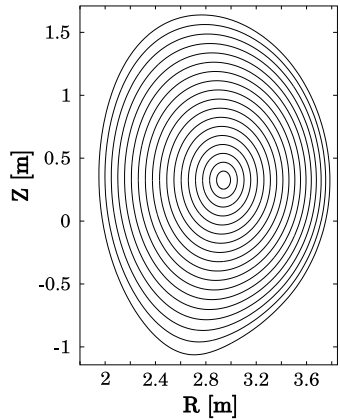}
    \caption{Cross section of the magnetic equilibrium, for JET discharge
             \#67730.}
    \label{fig:surf}
\end{figure}

The gyrokinetic simulations were performed in three different geometries:
$s-\alpha$, concentric circular~\cite{Lapillonne2009} and an MHD equilibrium
calculated for the JET-discharge \#67330 from EFIT data using the the TRACER
code (see~\cite{Xanthopolous2006} for details on the geometry implementation).
A cross~section of the experimental magnetic equilibrium is shown in
\Figref{fig:surf}.
For the parameters under consideration, $\alpha$ is fairly small
($\alpha=0.1026$), meaning that the $s-\alpha$ and circular model should be
near equivalent.
The $s-\alpha$ model, however, suffers from inconsistencies of the order
$\epsilon=a/R$ in its standard implementation~\cite{Xanthopolous2006,
Lapillonne2009, Burckel2010}.
Supplementary simulations using circular geometry were therefore performed, in
order to better differentiate between effects of magnetic geometry and effects
due to the inconsistent $s-\alpha$ metric.

    \subsection{Impurity transport}
    \label{sec:PF}
    For trace impurities, \Eqref{eq:transport} can be uniquely
written~\cite{Nordman2011} as a linear function of $\grad n_Z$, offset by a
convective velocity or ``pinch'' $V_Z$:
\begin{equation} \label{eq:transport_short}
    \Gamma_{Z} =
    -D_Z\grad n_Z + n_Z V_Z \Leftrightarrow \frac{R\Gamma_Z}{n_Z} =
    D_Z\frac{R}{L_{n_Z}} + RV_Z,
\end{equation}
\noindent where $D_Z$ is the impurity diffusion coefficient, and $V_Z$ is the
impurity convective velocity, which includes roto-diffusion in the gyrokinetic
treatment.
Both $D_Z$ and $V_Z$ are independent of $\grad n_Z$ in the trace impurity
limit~\cite{Angioni2006}.

In the core of a steady-state plasma with fuelling from the edge (i.e. in the
absence of particle sinks or sources in the core), the impurity flux $\Gamma_Z$
will go to zero.
Since the gradient is a measure of how peaked the impurity density profile is,
the gradient of zero impurity flux is referred to as the impurity peaking
factor ($PF$).
Inserting $R/L_{n_Z}=PF$ into the linearised \Eqref{eq:transport_short}, it can
be seen that the peaking factor quantifies the balance between convective and
diffusive impurity transport:\footnote{this definition is closely related to
the \emph{Péclet number}~\cite{Arter1995, Bakunin2003}}
\begin{equation}
    \label{eq:PF}
    PF=-\frac{R\,V_Z}{D_Z}.
\end{equation}
\noindent Specifically, the sign of the peaking factor is determined by the
sign of the pinch, meaning that $PF>0$ is indicative of a net inward impurity
pinch, giving a peaked impurity profile.
Conversely, if $PF<0$ the net impurity pinch is outward, leading to a hollow
impurity profile (also called \emph{flux reversal}).
For trace impurities, \Eqref{eq:transport_short} is linear in $R/L_{n_Z}$, and
the peaking factor is found by calculating the impurity particle flux
($\Gamma_Z$) for a range of $R/L_{n_Z}$ and solving \Eqref{eq:transport_short}
for zero flux.
For non-trace species, the $D_Z$ and $V_Z$ depend on $R/L_{n_Z}$, and the
peaking factor has to be found by explicitly seeking the gradient yielding zero
particle flux.

Much of the observed difference between the TE~and ITG~mode dominated cases can
be understood from the convective velocity $V_Z$ in \Eqref{eq:transport_short}.
To lowest order in $Z^{-1}$, the pinch contains two terms that depend on
$Z$~\cite{Angioni2006}:
\begin{itemize}
    \item thermal diffusion (thermopinch):
    \begin{itemize}
        \item $V_{\grad T_Z}\sim \frac{1}{Z}\frac{R}{L_{T_Z}}\omega_r$,
        \item inward for TE mode $\left(V_{\grad T_Z}<0\right)$,
              outward for ITG mode $\left(V_{\grad T_Z}>0\right)$,
    \end{itemize}
\item parallel impurity compression:
    \begin{itemize}
        \item $V_{\parallel,Z}\sim\frac{Z}{A_Z}k_\parallel^2 \sim
              \frac{Z}{A_Z q^2}\approx\frac{1}{2 q^2}$,
        \item outward for TE mode $\left(V_{\parallel_Z}>0\right)$,
              inward for ITG mode $\left(V_{\parallel_Z}<0\right)$.
    \end{itemize}
\end{itemize}
\noindent Here $1/k_\parallel$ represents the wavelength of the parallel
structure of the turbulence.
Due to the ballooning character of the modes considered, this is proportional
to the safety factor ($q$).
In addition, the convective velocity contains the curvature pinch, which to
lowest order is independent of the impurity charge~\cite{Nordman2011}.

The $Z$ dependence in the parallel impurity compression is expected to be weak,
since the mass number is approximately $A_Z \approx 2Z$ for an impurity species
with charge $Z$, and for the high $q$ considered, this term is expected to be
small compared to the other pinch contributions.
The thermodiffusive contribution, however, can dominate the transport for low
$Z$ impurities (such as the Helium ash).
For lower charge numbers, the second order correction to the thermal pinch
can become important:
\begin{equation}
    \label{eq:thermal}
    V_T \sim \left(\omega_r \frac{T_Z}{T_e Z} -
             \frac{7}{4} \left(\frac{T_Z}{T_e Z}\right)^2 \right)
             \frac{R}{L_{T_Z}},
\end{equation}
where the second order term is inward, independent of the mode direction, and
finite Larmor-radius effects have been neglected~\cite{Nilsson1990,
Weiland2000, Angioni2006, Nordman2011}.

The direction of the contributions to the pinch are governed mainly by the
considered mode's direction of rotation, which in the considered cases has a
different sign for TE~and ITG~modes~\cite{Weiland2000}.
Expanding the convective velocity in \eqref{eq:transport_short} into thermal
diffusion and ``pure convection'',
\begin{equation} \label{eq:transport}
    \Gamma_{Z} =
    -D_Z\grad n_Z + n_Z \left(D_{T,Z}\grad T_Z + V_{p,Z}\right),
\end{equation}
their relative contributions to the total peaking factor can be uniquely
determined for trace impurities, using the method described in
e.g.~\cite{Casson2010}.
Here $V_{p,Z}$ includes the parallel compression and curvature pinches.
In the presence of sheared toroidal rotation ($v_\phi$), these terms are joined by a
roto-diffusive term, which is proportional to $\Dd v_\phi /\Dd r$.

\section{Results and discussion}
\label{sec:results}
    \subsection{Background turbulence and transport}
    \label{sec:bgturbulence}
        \subsubsection{Linear eigenvalue spectra}
        \label{sec:eigenspectra}
\begin{figure}[tb]
    \centering
    \subfloat[eigenvalue spectra for different geometry models]
             {\includegraphics{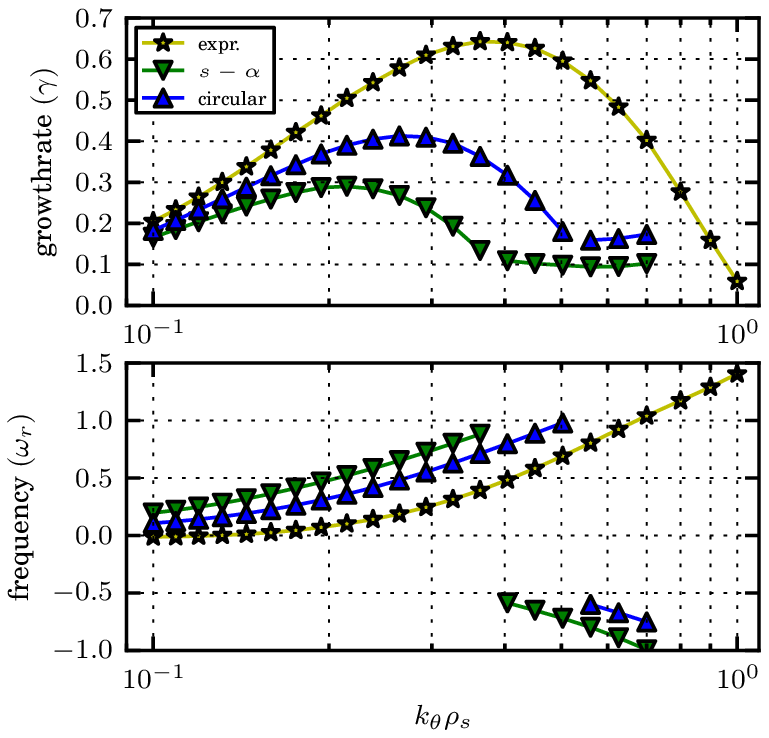}\label{fig:growthrates_geom}}~
    \subfloat[eigenvalue spectra in experimental geometry with added effects]
             {\includegraphics{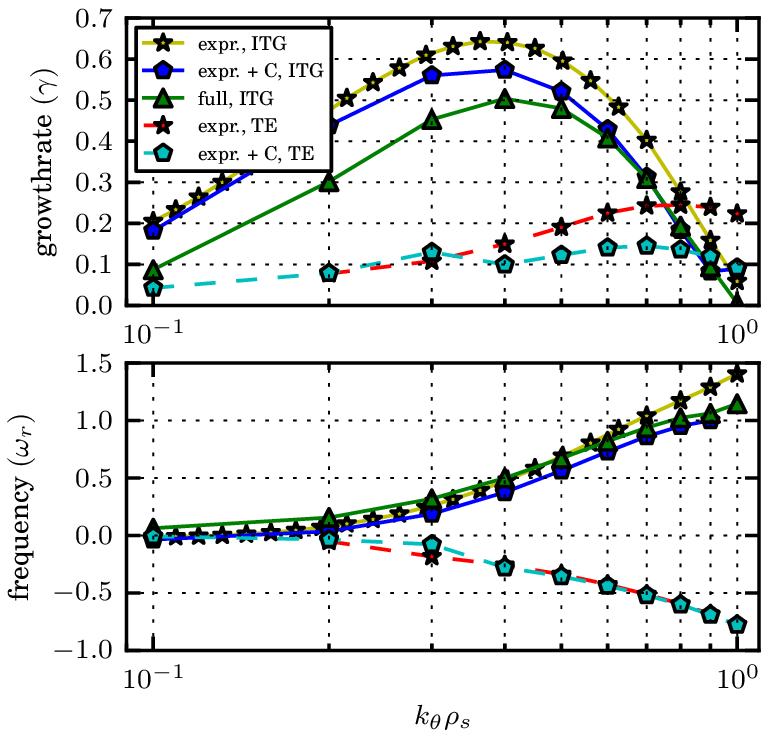}\label{fig:growthrates_fx}}
    \caption{(colour online)
             Growthrate spectra for $s-\alpha$, circular and experimental
             magnetic equilibrium for an ITG mode dominated case with JET-like
             parameters (\ref{fig:growthrates_geom}), and for the experimental
             equilibrium with added degrees of realism
             (\ref{fig:growthrates_fx}).
             For the second figure, \emph{expr.} represents full geometry case
             with no added effects, in \emph{expr.+C} a background of $2\%C$ was
             added, and in \emph{full} collisions are added. The last case
             corresponds to the ``full-scenario'' NL runs in
             \Figref{fig:PF_NL}.
             Parameters are presented in \Tabref{tab:parameters}.
            }
\label{fig:eigenspectra}
\end{figure}

Using GENE, the linear eigenvalues in the studied geometries were calculated at
mid-radius of JET L-mode discharge \#67330 for a range of different wave
numbers.
The resulting spectra are compared in~\Figref{fig:eigenspectra}.
The results in~\Figref{fig:growthrates_geom} show an overall destabilisation
and shift toward higher $k_\theta\rho_\text{s}$ when moving from $s-\alpha$ to
more realistic geometries, consistent with the results reported in
\cite{Lapillonne2009, Anderson2000}. 
For triangularity of the same order as the experimental value
($\delta=0.044$), the effect on the eigenvalues of $\delta$ was found to
be minute, and it is therefore surmised that the elongation is the main factor
behind the the observed results.

Adding more realism in the form of collisions and a $2\%$ carbon background, in
accordance with the ``full scenario'' considered in the nonlinear runs, both
have a stabilising effect, as can be seen in~\Figref{fig:growthrates_fx}.
The figure also shows that the stabilising effect is larger for the addition of
collisions than for the carbon background, in particular for lower values of
$k_\theta\rho_\text{s}$, where most of the transport normally occurs.
In the second figure, the spectra for the ITG mode is supplemented with those
of the sub-dominant TE mode, except in the full scenario, where it is completely
stabilised, with $\gamma < 0$ for the wave-numbers considered.

\begin{figure}
    \centering
    \includegraphics{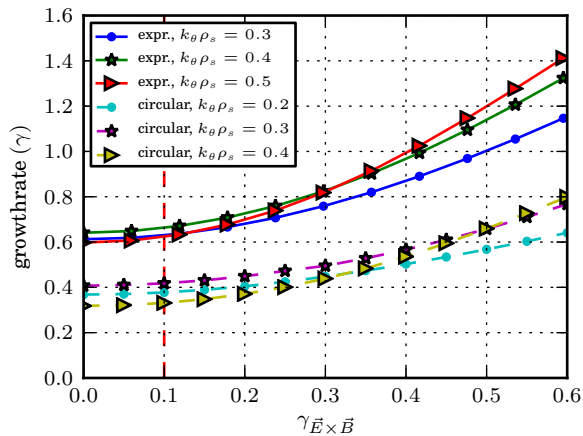}
    \caption{(colour online)
             Scaling of ITG growthrate with $\vec{E}\times\vec{B}$ shearing
             rate.
             The experimental shearing rate has been indicated (vertical dashed
             line).  }
    \label{fig:rot}
\end{figure}

The effects of sheared toroidal rotation on the ITG growth rate in the circular and
experimental equilibrium are shown in~\Figref{fig:rot}.
Both the stabilising perpendicular velocity shear and the destabilising
parallel shearing rate are included.
For the value of $q$ considered, the mode is destabilised with increasing
$\gamma_{\vec{E}\times\vec{B}}$.
This is because the stabilising component is proportional to $1/q$, wherefore the destabilising effect
dominates in the present case~\cite{Barnes2011}.
The shearing rate in the considered JET experiment (marked in the figure) is,
however, too small to have a significant impact on the mode growth, and is left
out of the nonlinear treatment below.

        \subsubsection{Nonlinear transport and fluctuation levels}
        \label{sec:spectraces}
\begin{figure}
    \centering
    \subfloat[timeseries of particle fluxes ($\Gamma$) with mean flux indicated]
             {\includegraphics{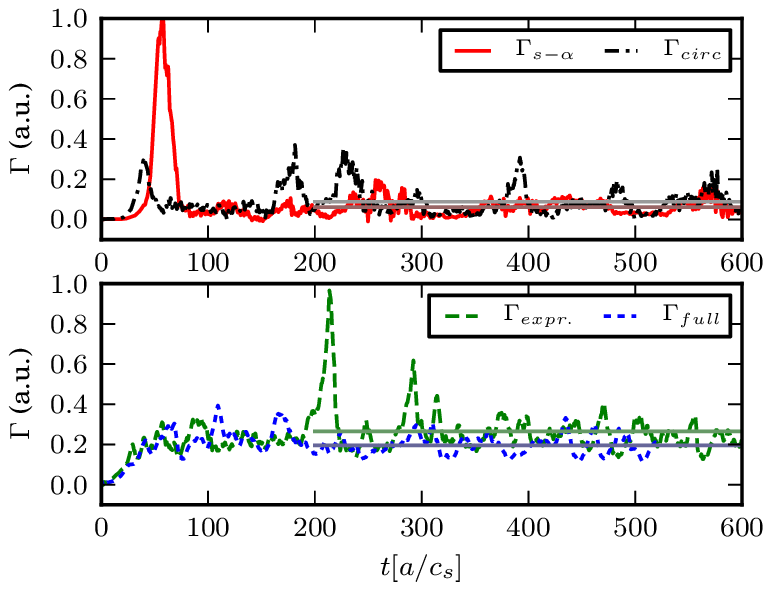} \label{fig:Gammatime}}~
    \subfloat[timeseries of heat fluxes ($Q$) with mean flux indicated]
             {\includegraphics{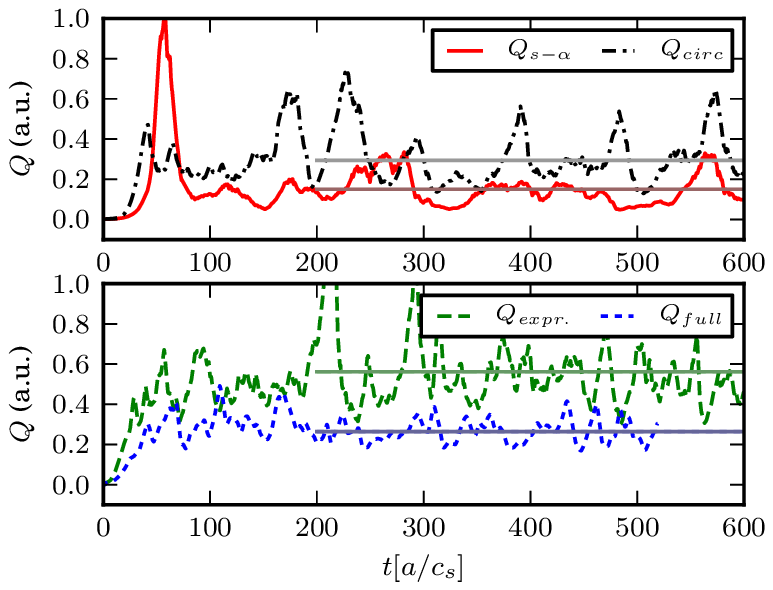} \label{fig:Qtime}}

    \subfloat[spectra of particle fluxes ($\Gamma$)]
             {\includegraphics{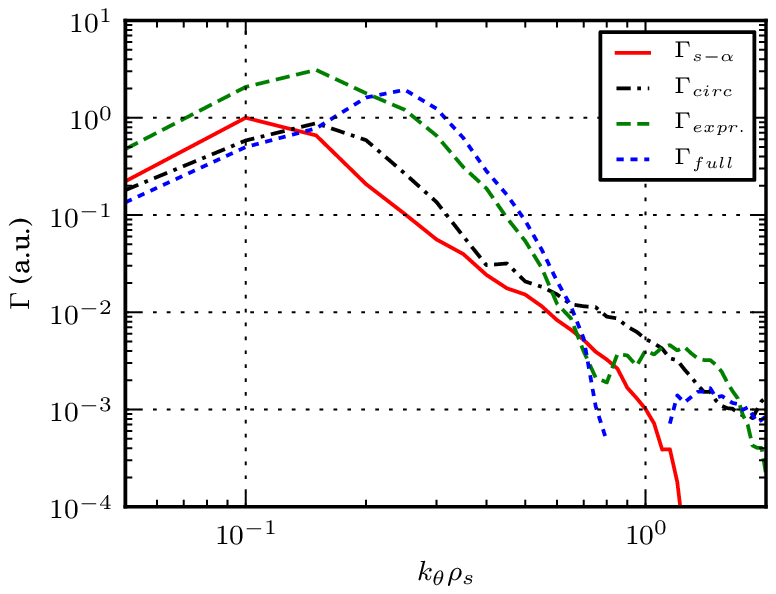} \label{fig:Gammalog}}~
    \subfloat[spectra of heat fluxes ($Q$)]
             {\includegraphics{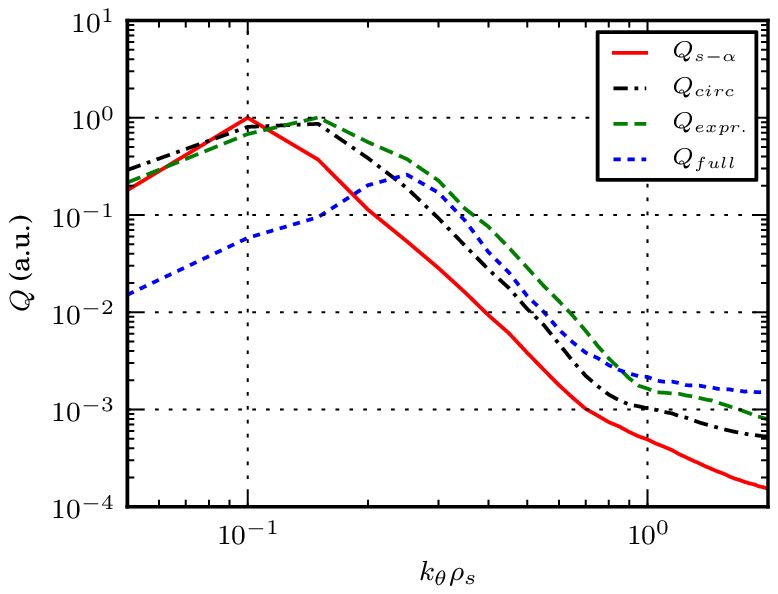} \label{fig:Qlog}}
    \caption{(colour online)
             Timeseries and spectra of particle ($\Gamma$) and heat ($Q$)
             fluxes for nonlinear GENE simulations for $s-\alpha$ and circular
             geometry, the experimental equilibrium, and the fully realistic
             case, with parameters as in \Tabref{tab:parameters}.
             Spectra show average over radial wavenumbers.
             The timeseries and spectra have been normalised to the maximum of
             the corresponding $s-\alpha$ entry.
             The spectra and time series show a decrease in amplitude
             consistent with that expected from the linear eigenvalue spectra
             in~\Figref{fig:eigenspectra}.
        }
     \label{fig:fluxes}
 \end{figure}

By comparing the transport levels for the different geometries, the difference
in the turbulence can be assessed qualitatively.
Time series and wave-number spectra of heat and particle flux ($Q$ and $\Gamma$
respectively) for the main ions were produced from nonlinear GENE simulations.
The results are presented in \Figref{fig:fluxes}.

As can be seen in~figures~\ref{fig:Gammatime} and~\ref{fig:Qtime}, the
turbulent fluctuation level is increased when moving from $s-\alpha$ to
realistic magnetic geometry.
The same trend holds true for the spectra in figures~\ref{fig:Gammalog}
and~\ref{fig:Qlog}.
Simulations using circular geometry~\cite{Xanthopolous2006, Lapillonne2009}
place both the spectra and time series between the $s-\alpha$ and
experimental case without collisions.
This is consistent with the linear eigenvalues in \Figref{fig:eigenspectra},
and with the results reported in \cite{Lapillonne2009}. 
With the addition of collisions and background carbon, the fluctuation
levels are brought down, due to a general reduction of the turbulence, as seen
in the the eigenvalue spectra \Figref{fig:growthrates_fx}.
The particle flux ($\Gamma$) is more sensitive to collisions than the heat
flux ($Q$), and therefore retains a higher fluctuation level, despite the lower
growthrates.

In the case with full realism, including collisions and carbon background, the
fluctuation levels are lowered to an intermediate level, consistent with higher
dissipation from collisions.
This is also consistent with the linear eigenvalues presented
in~\Figref{fig:eigenspectra}, in which the growthrates for the ``full
scenario'' have an intermediate value between the circular (or $s-\alpha$) and
the case where only effects of realistic geometry is considered.
As can be seen in figures~\ref{fig:Gammatime} and~\ref{fig:Qtime}, the full
case was simulated for a shorter time span, due to its larger computational
cost, mainly due to the inclusion of collisions, but also in part due to the
$2\%$C background; see Section~\ref{sec:gyro}.

We note that the global confinement time in both L-mode and H-mode plasmas are
scaling favourably with elongation, opposite to the trend found here for the
core transport.
This is likely a result of edge physics not included in the present
study~\cite{IPB1}.

        \label{sec:zonal}
In addition, the zonal flow activity was investigated for each case.
However, no significant differences were observed between the considered
cases.
The NL results can thus be explained qualitatively by the linear physics.

    \subsection{Impurity peaking factors}
    \label{sec:imptransport}
        \label{sec:imppeaking}

Simulations of impurity transport using a \emph{JET}-like experimental magnetic
equilibrium have been compared to simulations with $s-\alpha$ and circular
geometry, using the gyrokinetic code GENE.

\begin{figure}
    \centering
    \includegraphics{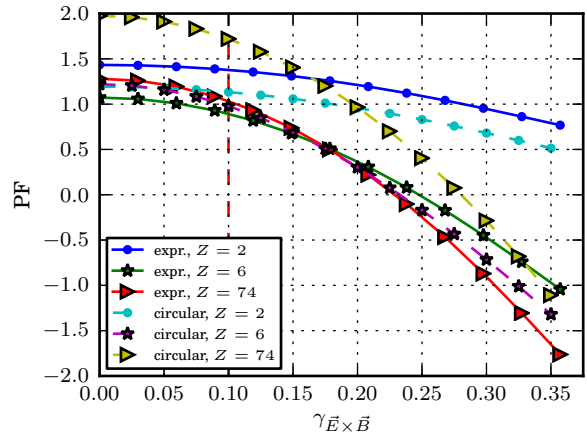}
    \caption{(colour online)
             Scaling of PF with $\vec{E}\times\vec{B}$ shearing rate for
             different impurity species in circular and shaped equilibrium for
             wavenumbers near the peak of the corresponding growthrate spectra
             ($k_\theta\rho_s=0.3$ and $k_\theta\rho_s=0.4$ respectively).
             The experimental shearing rate has been indicated (vertical dashed
             line).  }
    \label{fig:PF_rot}
\end{figure}

First, the effects sheared toroidal rotation in the different geometries is
investigated The effects on the impurity peaking factors are displayed in
\Figref{fig:PF_rot}.
The peaking factors for several impurity species are shown as a function of
$\gamma_{\vec{E}\times\vec{B}}$ in the experimental equilibrium.
As observed, the peaking factors are reduced by the sheared rotation.
This leads to a reversal of the impurity pinch
for $\gamma_{\vec{E}\times\vec{B}}\gtrsim 0.23$ for all but the lightest
impurities ($Z\lesssim 4$) in the experimental equilibrium.
In the circular geometry the trends are the same, but the flux reversal for
Tungsten ($Z=74$) is shifted to higher rotational sharing rate
($\gamma_{\vec{E}\times\vec{B}}\gtrsim 0.28$).
This effect, due to roto-diffusion, has been found to be a
critical ingredient to include in order to reproduce the Boron profile in
ASDEX~U~\cite{Angioni2011}.
For the considered discharge, however, the shearing rate is small and hence it
is neglected in the non-linear analysis.

\begin{figure}[tb]
    \centering
    \subfloat[comparison of $s-\alpha$, circular and experimental geometries for
              QL GENE]{
        \includegraphics{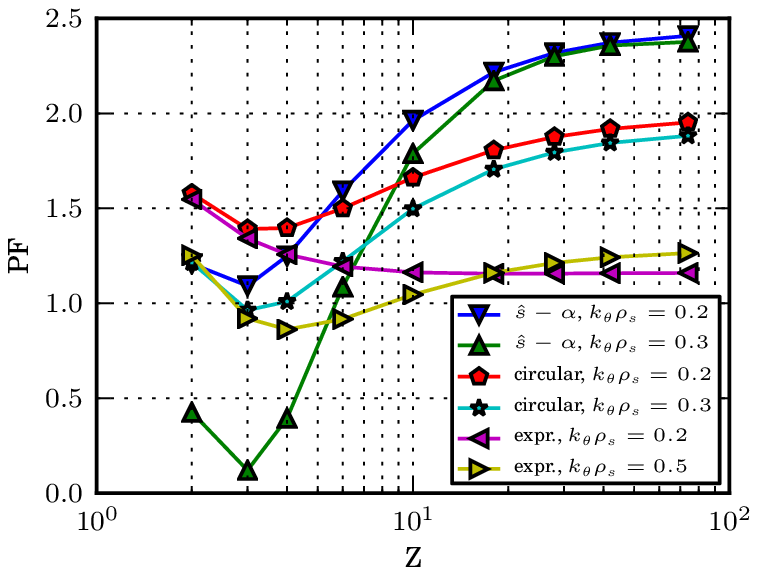} \label{fig:PF_QL_geom}}

    \subfloat[comparison of NL GENE base case and full case (\emph{expr.}
              and~\emph{full} in \Figref{fig:eigenspectra}) with results for $s-\alpha$
              geometry from~\cite{Nordman2011}; error-bars indicate a standard error
              of $\pm\sigma$.]{
        \includegraphics{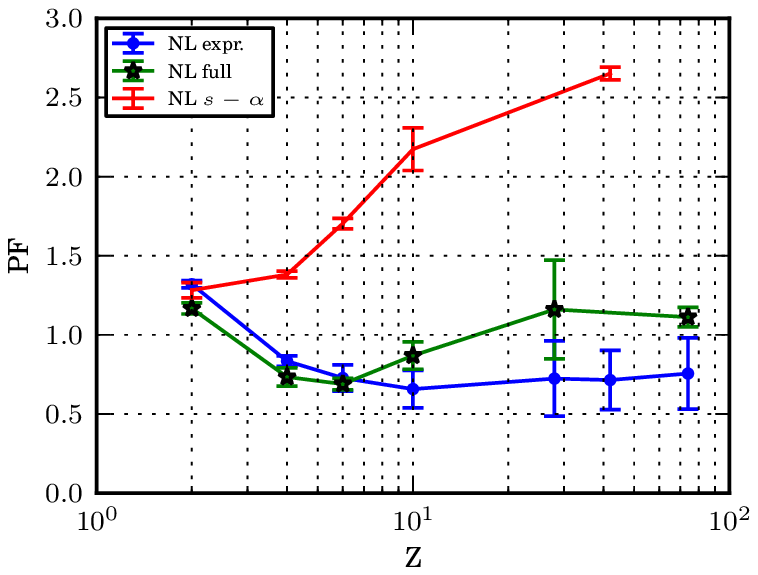}
        \label{fig:PF_NL}}~
    \subfloat[effects of added realism on QL $Z$-scaling for $k_\theta\rho_s=0.4$]{
        \includegraphics{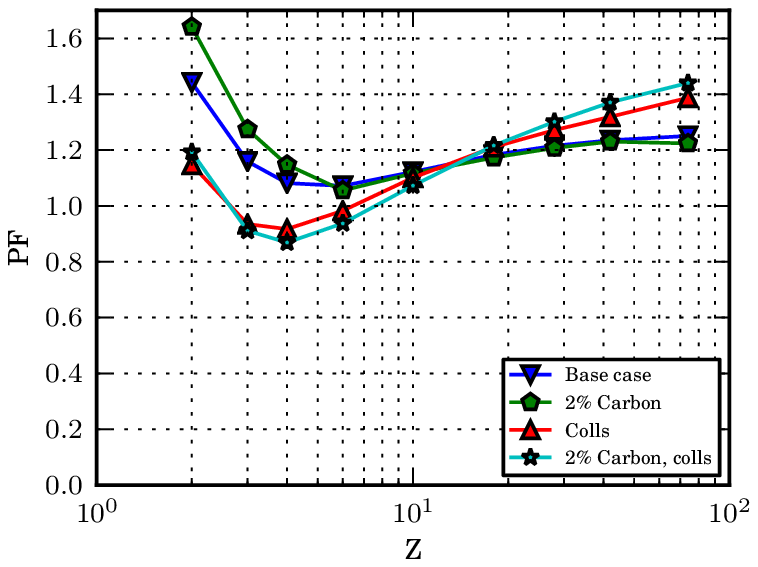} \label{fig:PFeffects}}
    \caption{(colour online)
             Scalings of impurity peaking factor~($PF$) with impurity charge
             number~($Z$) from JET discharge \#67730 for $s-\alpha$, circular
             and experimental equilibrium.
             Parameters as in \Tabref{tab:parameters}.
            }
    \label{fig:Z}
\end{figure}

Next, the effects of the geometry on the scaling with impurity charge are
investigated.
In \Figref{fig:PF_QL_geom} the scaling of the impurity peaking factor ($PF$)
with impurity charge ($Z$) is shown, for $s-\alpha$, circular and experimental
geometry.
Non-linear (NL) gyrokinetic results are compared in
\Figref{fig:PF_NL} for added degrees of realism.
Error-bars in \Figref{fig:PF_NL} correspond to a conservative standard
error estimate of $\pm\sigma$.
This was calculated from the NL flux data, where an effective sample size was
gauged for the time series and used to estimate the error for the mean flux.
This estimate was then propagated through to an error estimate for the peaking
factor.
The effects on the $PF$-scaling of adding collisions and $2\%$C background,
consistent with the considered JET~discharge, are shown in
\Figref{fig:PFeffects}.


For moderate to high impurity charge, weaker scalings of $PF$ with $Z$ were
consistently observed when departing from the $s-\alpha$ equilibrium, and the
level at which $PF$ saturates for high $Z$ was reduced in the experimental
geometry case.
The lower $PF$ levels compared to $s-\alpha$ geometry can be attributed to a
reduction of the convective pinch due to effects of shaping, but also due to
effects of finite inverse aspect ratio ($\epsilon$) effects, as discussed in
Section~\ref{sec:profiles}.
To separate the effects of shaping from the inconsistent $\epsilon$ effects in
the $s-\alpha$ model, simulations using GENE's circular
geometry~\cite{Lapillonne2009} were performed.
Since $\alpha \ll 1$ in the studied discharge, the difference between
the results using circular and experimental equilibrium in
\Figref{fig:PF_QL_geom} can attributed to shaping effects, mainly due to
elongation.
Further, a large increase in $PF$ was observed for low $Z$ impurities, most
notably He impurities, in circular and experimental geometry.
This is due to a reduction of the outward thermopinch in the more realistic
equilibrium models, as discussed in Section~\ref{sec:pinches}.


In~\Figref{fig:PF_NL} non-linear results using $s-\alpha$ geometry
from~\cite{Nordman2011} are compared with NL gyrokinetic simulations using the
experimental equilibrium for two different sets of parameters.
The first NL set of runs only uses the experimental geometry, whereas
the full case also includes a carbon background and collisions, but neglects
effects of sheared toroidal rotation (\emph{full} in
\Figref{fig:eigenspectra}).
With the introduction of these two effects, $PF$ is lowered for low $Z$ and
increased for high $Z$.
As seen in the quasilinear runs, the NL GK results are lower than those
predicted by the $s-\alpha$ model, though a qualitative agreement is reached
in the full scenario, particularly for high $Z$.
When comparing QL and NL results, the former show a more dramatic scaling than
the latter and the QL results tend to over-estimate $PF$ for high $Z$, as can
be seen in \Figref{fig:PFeffects}.
In all models there is, as expected, a saturation of $PF$ in the high $Z$
limit~\cite{Angioni2006, Nordman2011, Skyman2011a}, and the observed NL and QL
impurity pinches qualitatively agree with the results in~\cite{Nordman2011,
Skyman2011a}.

Figure~\ref{fig:PFeffects} shows the effect of adding collisions and a $2\%$
carbon background to the $PF$ scaling with $Z$, using QL gyrokinetics.
The ``base case'' corresponds to the experimental MHD equilibrium, without any
added effects (\emph{expr.} in \Figref{fig:eigenspectra}).
In all cases $k_\theta \rho_\text{s}=0.4$ was used, since that is the
approximate value for the maximum linear growth rate, as shown in
\Figref{fig:eigenspectra}.
As with the eigenvalue spectra (\Figref{fig:growthrates_fx}), it is clear that
the addition of collisions to the model has a larger impact than the addition
of the $2\%$C background.
The addition of the carbon background slightly raises $PF$ for low $Z$, but the
addition of collisions lowers it considerably, as well as raising $PF$ for high
charge numbers, as seen for the NL runs.
These results are consistent with previous observations of the importance of
impurity--main ion collisions for core impurity transport~\cite{Angioni2005,
Moradi2009}.
We note that, although a reduction is seen in the peaking for carbon impurities
in the realistic case, the flat or hollow profile seen in experiments is not
reproduced.
Though a further reduction is expected from roto-diffusion, nonlinear
simulations of the full scenario with sheared toroidal rotation still show an
inward nett transport of the background carbon.

    \subsection{Contributions to the impurity pinch}
        \label{sec:pinches}
In order to gain more insight into the results presented in the last
section, the contributions to the peaking factor were calculated using the
method outlined in~\cite{Casson2010}.

\begin{figure}
    \centering
    \subfloat[contributions to the pinch for circular equilibrium]{
        \includegraphics{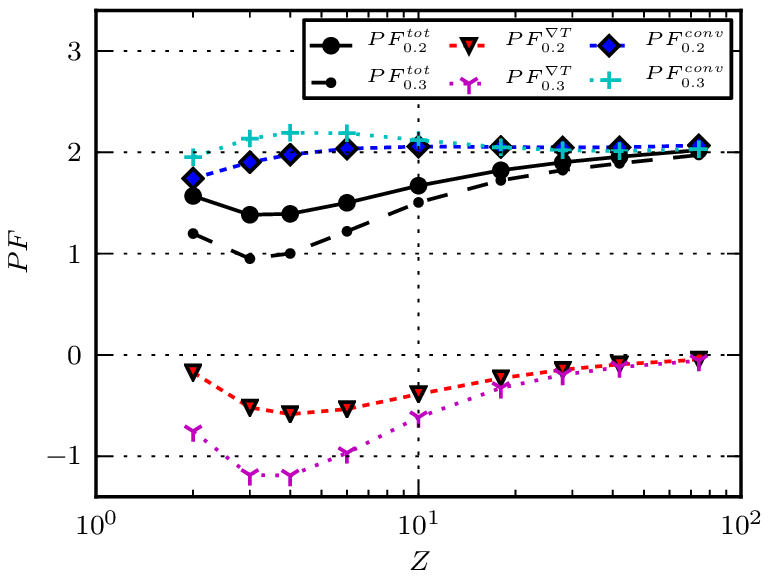} \label{fig:PF_contrib_circ}}~
    \subfloat[contributions to the pinch for experimental equilibrium]{
        \includegraphics{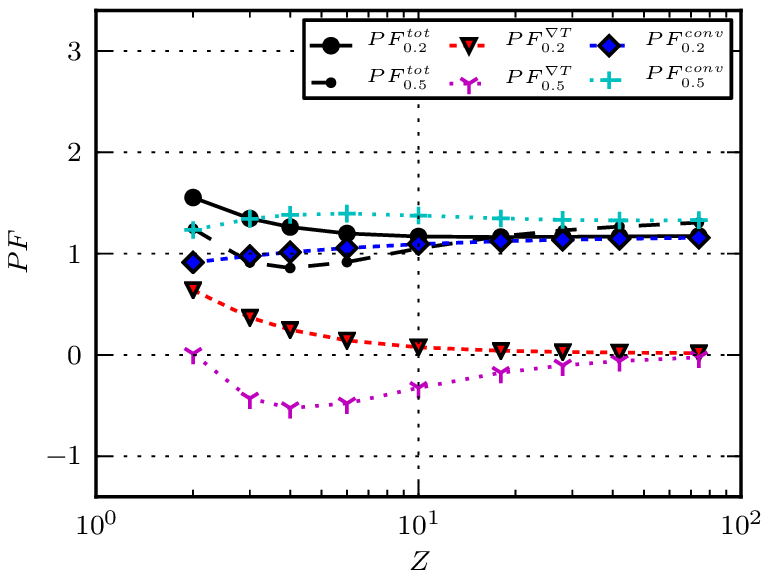} \label{fig:PF_contrib_full_rot}}
    \caption{(colour online)
        Contributions to the peaking factor
        ($PF_{k_\theta\rho_s}^\text{tot}$) from thermodiffusion
        ($PF_{k_\theta\rho_s}^{\grad T}$) and pure convection
        ($PF_{k_\theta\rho_s}^\text{conv}$) as a function of impurity charge
        ($Z$).
        Results from QL GENE for wavenumbers as in \Figref{fig:Z}, with and
        without effects of shaping ($\kappa=1.0$, $\kappa=1.37$).
        }
    \label{fig:contribs}
\end{figure}

Figure~\ref{fig:contribs} shows the effect of shaping on the thermodiffusive
and convective contributions to the impurity peaking factor
(\Eqref{eq:transport}).
It is seen, that the increase in peaking factor observed for lighter impurities
is mainly due to the thermopinch, where the inward second order contribution to
the thermal pinch comes into dominance over the outward term
(\Eqref{eq:thermal}).
This effect becomes more pronounced for lower wave-numbers -- especially in the
shaped equilibrium -- due to the lower mode frequency
(\Figref{fig:eigenspectra}).
For higher $Z$, however, the peaking is determined by the balance of the inward
convective pinch and the diffusion.
This contribution is only weakly dependent on the impurity charge and
wave-number, and is reduced substantially in the shaped equilibrium.

\begin{figure}
    \centering
    \includegraphics{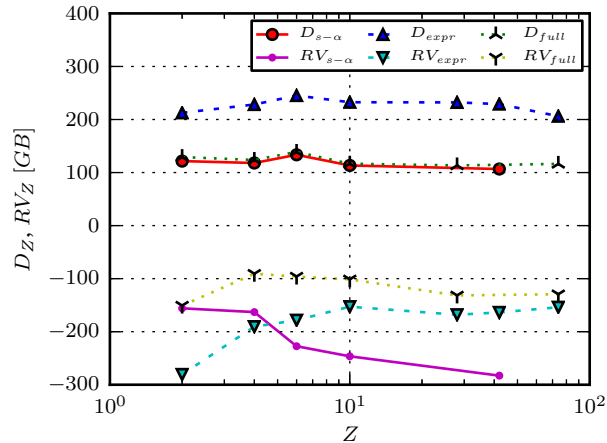}
    \caption{(colour online)
        Scaling of the impurity diffusivity ($D_Z$) and total pinch
        ($RV_Z$) with impurity charge for $s-\alpha$ geometry and experimental
        geometry, with and without collisions and $2\%$C background.
        Data from NL~GENE simulations with standard errors $~10$--$20\%$
        (omitted for clarity).
        }
    \label{fig:DRV}
\end{figure}

In
\Figref{fig:DRV}
diffusivities and pinches obtained from NL GENE are compared with with data
from~\cite{Nordman2011}.
From these results we note that, in contrast to the diffusivity ($D_Z$),
the convective velocity ($|V_Z|$) is lower in the realistic equilibrium as
compared to the $s-\alpha$ case, despite the corresponding increase in
fluctuation levels (\Figref{fig:Qtime}).
This confirms the conclusion that the main reason for the reduced peaking
factors obtained in the experimental equilibrium is a reduction of the
convective velocity.
When comparing the collisionless shaped equilibrium with the full scenario, we
note that the diffusivity is decreased, as expected from the observed decrease
in fluctuation levels with the addition of collisions.
For high $Z$, the reduction in convective velocity is similar in both cases,
compared to the $s-\alpha$ equilibrium, leading to higher peaking factors in
the full scenario.

When comparing the scalings for low ($k_\theta\rho_s=0.2$) and high ($k_\theta\rho_s=0.5$) wavenumbers in
the case with shaping, it is seen that the thermal diffusion is dominated
completely by the higher order terms ($\sim Z^{-2}$) in the low wavenumber
case, turning from mostly outward to inward.
In the full case, the NL spectrum was seen to shift toward higher
wavenumbers, compared to the case with shaping but without collisions and
carbon background (\Figref{fig:fluxes}).
The difference between the non-linear gyrokinetic results in \Figref{fig:PF_NL}
is therefore consistent with the interpretation that higher order terms
dominate the thermal diffusion for lower wave numbers, thereby determining the
shape of the peaking factor scaling.

\section{Conclusions}
\label{sec:conclusions}
The effects of the choice of equilibrium model on impurity transport in ITG/TE
mode driven turbulence were studied using gyrokinetic simulations.
Results were obtained and contrasted for the experimental MHD equilibrium and
the simpler $s-\alpha$ and concentric circular geometries.
These results were extended by adding degrees of realism in the physics
description, such as a 2\% carbon background, collisions, and sheared
toridal rotation.
The gyrokinetic results, obtained with the GENE code in both quasi- and
non-linear mode, were compared with results from a previous study, as well as
with expectations from a dedicated impurity injection L-mode discharge at JET.
It is found that the different equilibria give qualitatively similar results,
but with significant quantitative differences.

Linearly, a destabilisation and shift of the growthrate spectrum to higher wave
numbers was seen when departing from the simpler geometries, mainly due to
elongation.
This resulted in larger heat and particle transport, as seen in the nonlinear
simulations.
The addition of collisions was seen to stabilise the spectrum and reduce
transport.

The effect of sheared rotation on the mode stability and the impurity transport
was studied.
A weak destabilisation of the ITG mode with $\vec{E}\times\vec{B}$ shear was
found.
However, this was seen to be a minor contribution for experimentally feasible
values of the rotational shearing rate for the considered discharge.

The impurity peaking factors ($PF$s), computed by finding local density
gradients corresponding to zero particle flux, were derived.
The peaking factor was observed to saturate at levels far below neo-classical
predictions for high impurity charge numbers.
The level of the saturation was considerably lowered when the experimental
equilibrium was used, typically by a factor of $\sim 2$.
Comparing the diffusivity and convective pinch velocity in the different
geometries, it was found that the reduction of $PF$ was mainly due to a
reduction of the convective pinch.

By decomposing the contributions to the peaking factor into thermodiffusion
and pure convection, it was seen that second order contributions to the thermal
diffusion became important for low impurity charge.
Whereas the first order term is proportional to $\omega_r$, and therefore
outward for ion temperature gradient modes, the second order contribution to
the pinch is always inward, and therefore leads to higher $PF$s.
This explains the more pronounced increase in peaking observed for low charge
numbers in the shaped equilibria, where the real frequency is lower.

Collisions were seen to affect the impurity peaking factors considerably,
lowering $PF$ for light impurities, while increasing it for heavy impurities.
It was seen, that the addition of collisions and background carbon shifted
the nonlinear spectra towards higher wavenumbers, assosciated with higher real
frequencies.
Therefore, the the difference between the nonlinear results in the experimental
equilibrium could be also understood through the second order term in the
thermopinch, which was concluded to dominate this pinch contribution for
the non-collisional case.

However, even combining the effect of collisions with the effect of geometry
and rotation in the most realistic simulations, the observed peaking
factor for carbon was still too large to explain the rather flat C profile
observed in the experiment.
One explanation for this descrepancy could lie in the sensitivity of the
thermopinch to main ion and impurity temperature, and to the gradients of
thereof.
A moderate variation in these may bring the background carbon peaking factor
down to zero, however, such a treatment is left for future studies.
We also note that the treatment of rotation employed in this study neglects
centrifugal and Coriolis forces, which may lower the impurity peaking further.


\section*{Acknowledgements}
\label{sec:acknowledgements}
The main simulations were performed on resources provided on the
Lindgren\footnote{See
\protect\url{http://www.pdc.kth.se/resources/computers/lindgren/} for details
on Lindgren} high performance computer, by the Swedish National Infrastructure
for Computing (SNIC) at Paralleldatorcentrum (PDC).
Additional computations were carried out on resources at Chalmers Centre for
Computational Science and Engineering (C3SE)\footnote{See
\protect\url{http://www.c3se.chalmers.se}}, also provided by SNIC.

The authors would like to thank F~Jenko, T~Görler, F~Merz, MJ~Püschel, D~Told,
and the rest of the GENE~team at IPP--Garching for their valuable support and
input.

\bibliographystyle{iopart-num}
\bibliography{fusion}

\end{document}